\documentclass{aa}  
\usepackage{graphicx}
\usepackage{txfonts}
\usepackage{natbib}
\usepackage{dcolumn}

\newcolumntype{.}{D{.}{.}{-1}}
\newcolumntype{;}{D{;}{.}{7}}
\bibpunct{(}{)}{;}{a}{}{,} 

\newcommand{\solm}{M$_{\odot}$\ }

\begin{document}

\authorrunning{Eckart et al.}
\titlerunning{SgrA* Flares}
\title{Modeling mm- to X-ray flare emission from SgrA*}
\subtitle{}
   \author{A. Eckart\inst{1,2}
	  \and
          F. K. Baganoff\inst{3}
          \and 
          M. R. Morris\inst{4}
          \and 
          D. Kunneriath\inst{1,2}
          \and
          M. Zamaninasab\inst{1,2}
          \and
          G. Witzel\inst{1}
          \and
          R. Sch\"odel\inst{9}
          \and
          M. Garc\'{\i}a-Mar\'{\i}n\inst{1}
          \and
          L. Meyer\inst{4}
          \and
          G.C. Bower \inst{5}
          \and
          D. Marrone \inst{6}
          \and
          M.W. Bautz\inst{3}
          \and
          W.N. Brandt\inst{7}
          \and
          G.P. Garmire\inst{7}
          \and
          G.R. Ricker\inst{3}
          \and
          C. Straubmeier\inst{1}
          \and
          D.A. Roberts  \inst{8}
          \and
          ~K. Muzic\inst{1,2}
          \and
          ~J. Mauerhan\inst{4} 
          \and
	  ~A. Zensus\inst{2,1}
          }
\offprints{A. Eckart (eckart@ph1.uni-koeln.de)}

   \institute{ I.Physikalisches Institut, Universit\"at zu K\"oln,
              Z\"ulpicher Str.77, 50937 K\"oln, Germany\\
              \email{eckart@ph1.uni-koeln.de}
         \and
             Max-Planck-Institut f\"ur Radioastronomie, 
             Auf dem H\"ugel 69, 53121 Bonn, Germany
         \and
             Center for Space Research, Massachusetts Institute of
              Technology, Cambridge, MA~02139-4307, USA\\
             \email{fkb@space.mit.edu}
         \and
             Department of Physics and Astronomy, University of 
             California Los Angeles, Los Angeles, CA~90095-1562, USA\\
             \email{morris@astro.ucla.edu}
         \and
             Department of Astronomy and Radio Astronomy Laboratory,
             University of California at Berkeley, Campbell Hall, 
             Berkeley, CA~94720, USA\\
             \email{gbower@astro.berkeley.edu}
         \and
             Harvard-Smithsonian
             Center for Astrophysics, Cambridge MA 02138, USA\\
             \email{dmarrone.cfa.harvard.edu}
         \and
             Department of Astronomy and Astrophysics, Pennsylvania
              State University, University Park, PA~16802-6305, USA
         \and
             Department of Physics and Astronomy, 
             Northwestern University, Evanston, IL 60208
         \and
             Instituto de Astrof\'isica de Andaluc\'ia,
	     CSIC, Camino Bajo de Hu\'etor 50, 18008 Granada, Spain 
             \email{rainer@iaa.es}
             }

\date{Received  / Accepted }

\abstract{We report on new modeling results based on the mm- to X-ray emission 
of the SgrA* counterpart associated with the
massive $\sim$4$\times$10$^6$\solm ~black hole at the Galactic Center.  
}{
We investigate the physical processes 
responsible for the variable emission from SgrA*.
}{
Our modeling is based on simultaneous observations  
carried out on 07 July, 2004, using the NACO adaptive
optics (AO) instrument at the European Southern Observatory's Very Large
Telescope\footnote{Based on observations at the Very Large Telescope
(VLT) of the European Southern Observatory (ESO) on Paranal in Chile;
Program: 271.B-5019(A).} and the ACIS-I instrument aboard the
\emph{Chandra X-ray Observatory} as well as the Submillimeter Array
SMA\footnote{The Submillimeter
Array is a joint project between the Smithsonian Astrophysical
Observatory and the Academia Sinica Institute of Astronomy and
Astrophysics, and is funded by the Smithsonian Institution and the
Academia Sinica.} on Mauna Kea, Hawaii,
and the Very Large Array\footnote{The VLA is operated by the 
National Radio Astronomy Observatory 
which is a facility of the National Science Foundation operated 
under cooperative agreement by Associated Universities, Inc.}
in New Mexico.  
}{
The observations revealed several flare events in all wavelength domains.
Here we show that the flare emission can be described with a combination of 
a synchrotron self-Compton (SSC) model followed by an adiabatic expansion of the source components.
The SSC emission at NIR and X-ray wavelengths involves up-scattered
sub-millimeter photons from a compact source component.
At the start of the flare, spectra of these components peak at frequencies 
between several 100~GHz and 2~THz.
The adiabatic expansion  then accounts for the
variable emission observed at sub-mm/mm wavelengths.
The derived physical quantities that describe the flare emission give a blob
expansion speed of v$_{exp} \sim 0.005$c, magnetic field of
B around 60~G or less and spectral indices of $\alpha$=0.8 to 1.4, 
corresponding to a particle spectral index p$\sim$2.6 to 3.8.
}{
A combined SSC and adiabatic expansion model can fully account for the 
observed flare flux densities and delay times covering the spectral range 
from the X-ray to the mm-radio domain.
The derived model parameters suggest that the adiabatic expansion 
takes place in source components that have a bulk motion larger 
than v$_{exp}$ or the expanding material contributes to a corona or disk, 
confined to the immediate surroundings of SgrA*.
}


   \titlerunning{Modeling flare emission from SgrA*}
   \authorrunning{Eckart, Baganoff, Morris, Kunneriath et al.}  
   \maketitle
%

\section{Introduction}
\label{section:Introduction}

Stellar motion and variable emission allow us to associate Sagittarius A* 
(SgrA*) at the center of the Milky Way with a super-massive black hole 
(Eckart \& Genzel 1996, Genzel et al. 1997, 2000, Ghez et al. 1998, 2000, 2003,
2005, Eckart et al. 2002, Sch\"odel et al. 2002, 2003, Eisenhauer 2003, 2005).  

Recent radio, and near-infrared through X-ray observations 
have detected flaring 
and polarized emission and give detailed insight into the 
physical emission mechanisms at work in SgrA*
(e.g. Baganoff et al. 2001, 2002, 2003, Eckart et al. 2003, 2004, 2006ab, 2008ab, 
Porquet et al. 2003, 2008, Goldwurm et al. 2003, Genzel et al. 2003, 
Ghez et al. 2004a, and Eisenhauer et al. 2005, Hornstein et al. 2007,
Yusef-Zadeh et al. 2006ab, 2007, 2008, Marrone et al. 2008).  

Variability at radio through sub-millimeter wavelengths has been studied
extensively, showing that variations occur on timescales from hours to
years (Wright \& Backer 1993,
Bower et al. 2002, 2003, 2004, 2005a, 2006, Herrnstein et al. 2004, Zhao et al. 2003, 2004,
Eckart et al. 2006a, Mauerhan et al. 2005, Yusef-Zadeh et al. 2007, 2008, 
Miyazaki et al. 2006, Marrone et al. 2008).
Several flares have provided evidence for decaying millimeter and
sub-millimeter emission following NIR/X-ray flares.
Simultaneous multi-wavelength observations indicate 
the presence of adiabatically expanding source components with 
a delay between the X-ray and sub-mm flares of about 100 minutes
(Eckart et al. 2006a, Yusef-Zadeh et al. 2008, Marrone et al. 2008).
The adiabatic expansion is also supported by the expected 
swing in polarization as 
indicated by the measurements of Yusef-Zadeh et al. (2008).
From modeling the mm-radio flares at individual frequencies
Yusef-Zadeh et al. (2007, 2008)
invoke expansion velocities in the range from $v_{exp}$=0.003-0.1c, 
This is also supported by the results of recent NIR/sub-mm 
observations in May 2008 using NACO at the VLT and
the LABOCA bolometer at the Atacama Pathfinder Experiment (APEX),
at 0.87~mm wavelength (345~GHz)
(Eckart et al. 2008b, Garc\'{\i}a-Mar\'{\i}n 2008 in prep.). 
Here we find an expansion speed of 0.005~c.
The speed is well below the asymptotic upper limit of c/$\sqrt{3}$
obtained for a system of relativistically interacting particles
(e.g. Bowers 1972)
expected in the vicinity of the super-massive 
black hole (Blandford \& McKee 1977).
It is also low compared to 
the expected orbital velocities that may be of the
order of 0.5~c close to the last stable orbit around the SMBH.
The low expansion velocities 
suggest that the expanding gas cannot escape from SgrA*
or must have a large bulk motion (Yusef-Zadeh et al. 2008,
Eckart et al. 2008b).

In order to investigate this question in more detail we revisited the
first observations of a flare with simultaneous coverage in the
NIR/X-ray and sub-mm/mm wavelength domain 
observed on July 07, 2004. Eckart et al. (2006a) showed 
that the observed amplitudes of the flux density variations are
generally consistent with adiabatic expansion of a
synchrotron self-absorbed source (van der Laan 1966).
Here we present a detailed time dependent model of the flare emission
from the X-ray to the short cm-wavelength domain.

For optically thin synchrotron emission we refer throughout this paper
to photon spectral indices ($\alpha$) using
the convention $S_\nu\propto\nu^{-\alpha}$
and to spectral indices ($p$) of electron power-law distributions
using $N(E)\propto E^{-p}$ with $p=\left(1+2\alpha\right)$.
The assumed distance to SgrA* is 8~kpc (Reid 1993), consistent 
with more recent results (e.g. Ghez et al. 2005, Eisenhauer et al. 2003).


\section{Observations and data reduction}
\label{section:Observations}

In 2004 from July 05 to 08
Sgr~A* was observed from the radio millimeter to the X-ray wavelength
domain. 
On July 07 a strong simultaneous NIR/X-ray flare event was observed 
immediately followed by simultaneous SMA and VLA observations.
Putting emphasis particularly on the NIR/X-ray data the 
details for the entire observing run have been analyzed in
Eckart et al. (2006a).
For completeness we give in the following a brief summary of the 
data acquisition and reduction for the X-ray, NIR and sub-mm/mm domain
with emphasis on the essentials important for the presented analysis.
The observational results are summarized in Table~\ref{flares}.

\begin{figure}
\centering
\includegraphics[width=8.5cm,angle=-00]{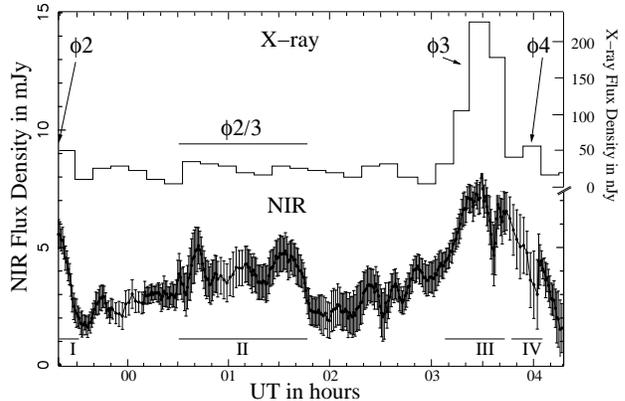}
\caption{\small  The X-ray and NIR 2.2 $\mu$m light curves obtained on July 07, 2004
(Eckart et al. 2006a). Here we plot the data with a UT time axis and 
separate flux density axes for the NIR (left) and X-ray (right) data. 
In addition to the flare nomenclature introduced in Eckart et al. (2006a)
we labeled the section of the X-ray light curve that corresponds to 
the NIR feature II as $\phi$2/3 as it is located between $\phi$2 and 
$\phi$3. 
}
\label{Fig:NIRXraydata}
\end{figure}

\begin{figure}
\centering
\includegraphics[width=9.0cm,angle=-00]{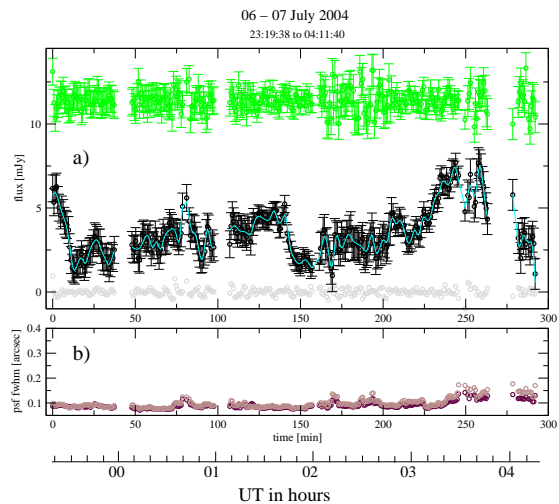}
\caption{\small  Results of the re-reduction of the NIR 2.2 $\mu$m 
data from July 07, 2004.
In panel a) the flux density scale is given in mJy. 
The top (green) data points represent the light curve of the nearby
flux star S1 used for flux density calibration
and the black data points with the blue interpolation line 
represent the SgrA* light curve.
The time axis is given in minutes offset from the start time 
at 23:19:38 UT on 6 July 2004 and in UT hours at the bottom.
In the bottom panel b) we show the FWHM of the PSF in the AO images 
that reflect the combination of AO correction and input seeing.
}
\label{Fig:reredflare}
\end{figure}

\begin{table}[!h]
\begin{center}
{\begin{small}
\begin{tabular}{ccccclccc} \hline
X-ray     & NIR & X-ray flux    &NIR K-band  & \\
 flare    & flare & density     &flux density& \\
 ID       &  ID   & (nJy)       & (mJy)      &  \\
\hline
$\phi$2   &  I  & 31$\pm$27   & $\ge$5.7   \\
$\phi$2/3 & II  & $<$20       & $\sim$3.0  \\
$\phi$3   &III  & 223$\pm$27  & 6$\pm$1.5  \\
$\phi$4   & IV  &  37$\pm$27  & 5$\pm$1.5  \\
\end{tabular}
\end{small}}
\end{center}
\caption{NIR/X-ray flare flux densities.
The peak flux densities of the flares detected in the individual 
wavelength bands are given.
The X-ray flares $\phi$2, $\phi$3 and $\phi$4 have been detected simultaneously in 
the NIR (Eckart et al. 2006a). 
Individual AO images for the NIR event II presented by Eckart et al. (2006a) 
as well as the newly reduced light-curve shown here in 
Fig.\ref{Fig:reredflare}
demonstrate that Sgr~A* clearly was in an ``on'' state.
\label{flares}}
\end{table}

\subsection{The NACO NIR adaptive optics observations}
\label{section:NACO}

Near-infrared (NIR) observations of the Galactic Center (GC) were
carried out with the NIR camera CONICA and the adaptive optics (AO)
module NAOS (briefly ``NACO'') at the ESO VLT unit telescope~4 on
Paranal, Chile, during the nights between 05 July and 
08 July 2004\footnote{Based on observations at the Very Large Telescope
(VLT) of the European Southern Observatory (ESO) on Paranal in Chile;
Programs: 073.B-0775 July 2004}
In all
observations, the infrared wavefront sensor of NAOS was used to lock
the AO loop on the NIR bright (K-band magnitude $\sim$6.5) supergiant
IRS~7, located about $5.6''$ north of Sgr~A*.  
All exposures were sky subtracted, flat-fielded, and corrected for
dead or bad pixels. 
In order to enhance the signal-to-noise ratio
of the imaging data, we created median images comprising 9
single exposures each. Subsequently, PSFs were extracted from these
images with \emph{StarFinder} (Diolaiti et al. 2000).
The images were deconvolved with the
Lucy-Richardson (LR) and linear Wiener filter (LIN) algorithms. Beam
restoration was carried out
with a Gaussian beam of FWHM corresponding to the 
final resolution at 2.2 \,$\mu$m
of 60~milli-arcseconds.
The flux densities of the sources were measured by aperture photometry with
circular apertures of 52~mas radius and corrected
for extinction, using $A_{K} = 2.8$.
Calibration of the photometry and
astrometry was done with the known fluxes and positions of
9 sources within $1.6"$ of Sgr~A*.

In Fig.~\ref{Fig:NIRXraydata} we show the July 07 NIR and X-ray data in 
comparison.
Four NIR flares (I - IV) can be identified.
In Fig.~11 of Eckart et al. (2006a) individual AO images correspond 
to separate points in time and include the flares discussed here.
These images demonstrate that even during the weak NIR flare feature II 
Sgr~A* clearly was in an ``on'' state and significantly weaker before and after.
For the flare feature II the NIR flux density excess is of the order of 3~mJy.

To re-assess the presence of the flare zone II we re-reduced the
NIR data and show the results in Fig.~\ref{Fig:reredflare}.
The re-reduction includes the following additional features:
1) We used sub-pixel shifting of the data applying the
   `jitter`-routine in ECLIPSE (Devillard 1997).
2) We subtracted a constant background based on a StarFinder analysis.
3) We rejected low quality images based on the number of stars detected by
   StarFinder.
4) We removed low level common trends ($\le$ 15\%) that became apparent in the 
   reference star data by applying a detrending routine 
   (by Nicolas Marchili, IMPRS, MPIfR).
   The generation of the trend involves binning and splining of the reference 
   star data and is similar to a procedure described in Villata et al. (2004).
   The resulting trend then is then removed from the SgrA* NIR light 
   curve as well.
5) For comparison we also plot the FWHM of the nearby reference stars 
   as an indication of the combined instantaneous NIR seeing and 
   the quality of the AO correction.

We find that within the uncertainties the result of the re-reduction is 
in very good agreement with the original 
data reduction used in Fig.~\ref{Fig:NIRXraydata}.
 The improved analysis shows that
 flare zone II is a reliably feature in the NIR light curve.

\subsection{The Chandra X-ray observations}
\label{section:Chandra}

In parallel to the NIR observations, SgrA* was observed with \emph{Chandra}
using the imaging array of the Advanced
CCD Imaging Spectrometer (ACIS-I; Weisskopf et al., 2002) for
two blocks of $\sim$50\,ks on 05--07 July 2004 (UT). 
We reduced and analyzed the data using CIAO v2.3\footnote{Chandra
Interactive Analysis of Observations (CIAO),
http://cxc.harvard.edu/ciao} software with Chandra CALDB
v2.22\footnote{http://cxc.harvard.edu/caldb}. 

We extracted counts within radii of 0.5\arcsec, 1.0\arcsec, and
1.5\arcsec\ around Sgr~A* in the 2--8 keV band.  Background counts
were extracted from an annulus around Sgr~A* with inner and outer
radii of 2\arcsec\ and 10\arcsec, respectively, excluding regions
around discrete sources and bright structures (Baganoff et al. 2003).  
The mean (total) count rates within the inner radius subdivided into 
the peak count rates during a flare and the corresponding intermediate
quiescent flux values
are listed in Table~4 in Eckart et al. (2006a).
The background rates have been
scaled to the area of the source region.  
The 1.0\arcsec\ aperture provides the best compromise between maximizing
source signal and rejecting background.
In Fig.~\ref{Fig:NIRXraydata} we have 
labeled the section of the X-ray light curve that corresponds to 
the NIR feature II as $\phi$2/3, as it is located between $\phi$2 and 
$\phi$3. 
For $\phi$2/3 the X-ray flux density excess above the quiescent 
bremsstrahlung component of SgrA* is below 20~nJy.

\begin{figure}
\centering
\includegraphics[width=8cm,angle=-00]{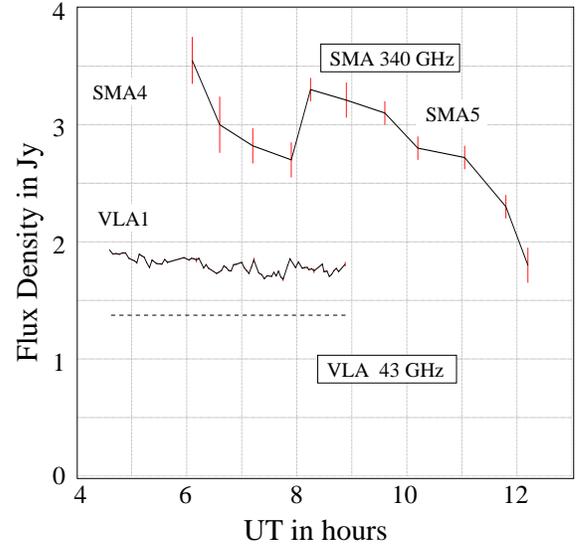}
\caption{\small
The 340~GHz SMA and 43~GHz VLA total intensity light curves from July 07.
The individual data points are connected by straight lines.
The July 07 VLA data represent the excess flux density compared to the 
mean of the July 06 and 08 VLA data. The constant flux density of 
about 1.4 Jy that has to be added to this excess is indicated by a dashed
line.
For further details see text and Eckart et al. (2006a).
}
\label{Fig:SMAVLAdata}
\end{figure}

\subsection{The SMA observations}
\label{section:SMA}
The sub-millimeter observations
were made with the Submillimeter Array\footnote{The Submillimeter
Array is a joint project between the Smithsonian Astrophysical
Observatory and the Academia Sinica Institute of Astronomy and
Astrophysics, and is funded by the Smithsonian Institution and the
Academia Sinica.} (SMA) on Mauna Kea, Hawaii (Ho, Moran, \& Lo 2004). 
The observations of SgrA* were made at 340~GHz (890~$\mu$m wavelength) 
for three consecutive nights, 05-07 July 2004 (UT),
at an angular resolution of 1\farcs5$\times$3\farcs0. 
Nearby quasars were used for phase and gain calibration. 
On both July 6 and
7 we obtained more than 6 hours of simultaneous X-ray/sub-millimeter
coverage with 340~GHz zenith opacities from 0.11 to 0.29 for July 5 to 
7, respectively. This is reflected in the larger time bins and scatter
in the later light curves.

The same 5 antennae with the best gain stability were
used to form light curves, resulting in a typical synthesized beam of
1\farcs5$\times$3\farcs0. 
The SgrA* data are phase self-calibrated after the
application of the quasar gains to remove short-timescale phase
variations, then imaged and cleaned. Finally, the flux density is
extracted from a point source fit at the center of the image, with the
error taken from the noise in the residual image. The overall flux
scale is set by observations of Neptune, with an uncertainty of
approximately 25\%.

We attribute a flux density value of $\sim$2.4~Jy as a constant or only 
slowly variable part or the light curve that may be due to more
extended source components. 
Since the final data point in the July 07 light curve is 
significantly below the minimum of $\sim$2.4~Jy that is usually 
obtained on SgrA* at 340~GHz 
(e.g. Yusef-Zadeh et al. 2008, Marrone et al. 2008)
and due to the steep drop in flux density towards the end of the 
observations at low elevations we did not consider this data 
point in our models of the light curve.

\subsection{The VLA 7~mm observations}
\label{section:VLA}

The Very Large Array (VLA) 
observed Sgr A* for $\sim 5$ hours on 6, 7 and 8 July 2004
at 43~GHz (7~mm wavelength). 
Observations covered roughly the UT time range 04:40 to 09:00,
which is a subset of the Chandra observing time on 6 and 7 July.  
Observations on 7 July immediately followed the VLT NIR observations.
The VLA was in D configuration and achieved a resolution of $2.5 \times 0.9$
arcsec at the observing wavelength of 0.7 cm.  
The absolute amplitude calibration was set by observations of 3C 286.  Flux densities
were determined for Sgr A* and J1744-312 through fitting of visibilities
at $(u,v)$ distances greater than 50 $k\lambda$ in order to remove
contamination from extended structure in the Galactic Center.

In Fig.~\ref{Fig:SMAVLAdata}
we show the 340~GHz SMA and 43~GHz VLA total intensity light curves from July 07.
Here the July 07 VLA light curve was calculated as the difference between
the mean flux density data at the same interferometer hour angle obtained
on July 06 and 08 (see Fig.~8 in Eckart et al. 2006a).
The minimum compact flux density of $\sim$1.4~Jy 
obtained between 7 and 8 hours UT has to be added to the resulting 
excess flux density in order to derive a complete 43~GHz 
light curve of SgrA* obtained in the VLA~~~D configuration 
at an angular resolution of 2\farcs5$\times$0\farcs9. 

\section{Radiation mechanisms}
\label{section:Mechanisms}

Due to the short flare duration the flare emission very likely originates 
from compact source components.
The simultaneous X-ray/NIR flare detections of the SgrA* counterpart
implies that the same population of electrons is responsible for 
both the IR and the X-ray emission (e.g. Eckart et al. 2004).
The spectral energy distribution of SgrA* is currently explained by models
that invoke radiatively inefficient accretion flow processes (RIAFs: Quataert
2003, Yuan et al. 2002, Yuan, Quataert, \& Narayan 2003, 2004,
including advection dominated accretion flows (ADAF): Narayan et
al. 1995, convection dominated accretion flows (CDAF): Ball et
al. 2001, Quataert \& Gruzinov 2000, Narayan et al. 2002, Igumenshchev
2002, advection-dominated inflow-outflow solution (ADIOS): Blandford
\& Begelman 1999; see also Ballantyne, \"Ozel, Psaltis 2007, Le \& Becker 2005), 
jet models (Markoff et al. 2001, see also Markoff 2005), and Bondi-Hoyle
models (Melia \& Falcke 2001). 
Also combinations of models such as an accretion flow 
plus an outflow in the form of a jet are considered 
(e.g. Yuan, Markoff, Falcke 2002).

\begin{figure}
\centering
\includegraphics[width=8.5cm,angle=-00]{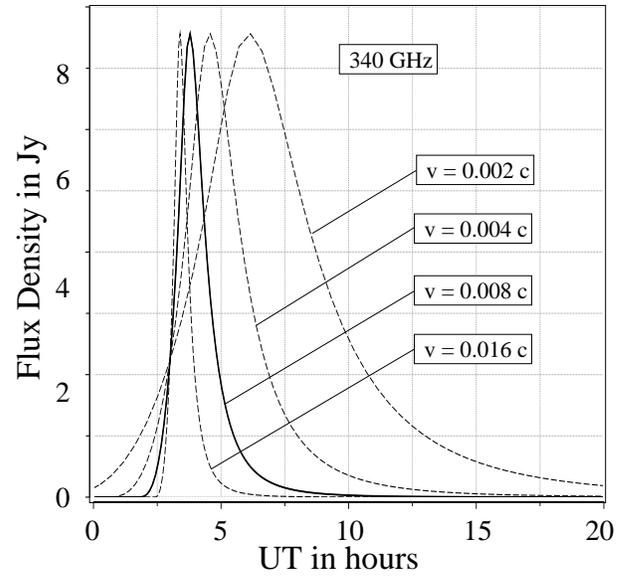}
\caption{\small  A comparison of a 340~GHz light curve 
calculated with adiabatic expansion velocities that differ by 
factors of two.
}
\label{Fig:speed}
\end{figure}

\begin{figure}
\centering
\includegraphics[width=8.5cm,angle=-00]{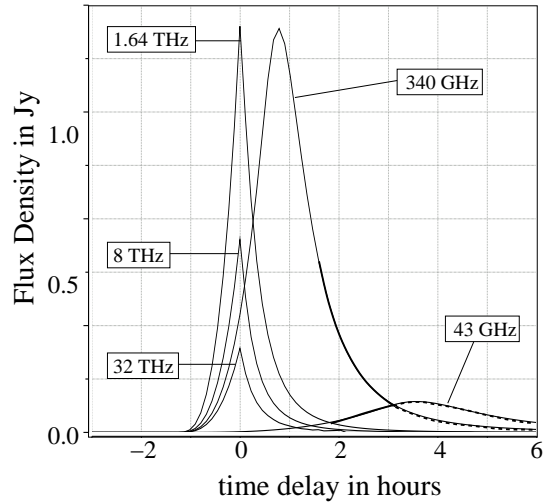}
\caption{\small  
The adiabatic expansion of a single source component
with a peak flux density at 1.64~THz of 10~Jy,
start time at 0~hours, and
constant expansion velocity of 0.08~c.
The 1.64~THz, 8~THz and 32~THz light curves have been scaled down by a factor of 5.2.
}
\label{Fig:single}
\end{figure}

\subsection{Adiabatically expanding source components}
\label{section:Adiabatic}

To model the sub-mm/mm light curves we assume an expanding uniform blob of 
relativistic electrons 
with an energy spectrum $n(E) \propto E^{-p}$
threaded by a magnetic field.  As the blob expands, the magnetic 
field declines  with increasing  blob radius as $R^{-2}$, the energy of relativistic 
particles as $R^{-1}$ and the density of particles as $R^{-3}$
(van der Laan 1966). 
The synchrotron optical depth 
at frequency $\nu$ then scales as 
\begin{equation}
    \tau_\nu = \tau_0 \left(\frac{\nu}{\nu_0}\right)^{-(p+4)/2}
    \left(\frac{R}{R_0}\right)^{-(2p+3)}
    \label{eq:taunu}
\end{equation}
and the flux density scales as
\begin{equation}
    S_\nu = S_0 \left(\frac{\nu}{\nu_0}\right)^{5/2} 
    \left(\frac{R}{R_0}\right)^3 
    \frac{1-\exp(-\tau_\nu)}{1-\exp(-\tau_0)}~~~.
    \label{eq:Snu}
\end{equation}
Since the goal is to combine the description of an adiabatically expanding
cloud with a synchrotron self-Compton formalism we use the definition of
$\tau_0$ as the optical depth corresponding to the frequency at which the 
flux density is a maximum (van der Laan 1966) rather than the definition of
$\tau_0$ as the optical depth at which the flux density for any
particular frequency peaks (Yusef-Zadeh et al.\ 2006b).  
Therefore $\tau_0$ depends only on $p$ through the condition
\begin{equation}
    e^{\tau_0} - \tau_0(p + 4)/5 - 1 = 0
    \label{eq:tau0}
\end{equation}
and ranges from 0 to 0.65 as $p$ ranges from 1 to 3.
Thus given the particle energy spectral index $p$ and
the peak flux $S_0$ in the light curve at some frequency $\nu_0$, this
model predicts the variation in flux density at any other frequency
as a function of the expansion factor $(R/R_0)$.

A model for $R(t)$ is required to convert the dependence on
radius to time: we adopt a simple linear expansion at
constant expansion speed v$_{exp}$, so that $R-R_0 = $v$_{exp}\,(t-t_0)$.
Here we assume that the source component is decoupled from energy input
and is freely expanding, i.e. neither
accelerations nor decelerations of the expansion are dominant
(see end of section \ref{section:bulkmotion}).
A possible magnetic confinement of the spot 
will be described in the model as a low expansion speed, i.e. it is 
contained in the model of $R(t)$.
For $t \le t_0$ we have made the assumption that the source has an optical 
depth that equals its frequency dependent initial value 
$\tau_{\nu}$ at $R = R_0$.
So in the optically thin part of the source spectrum the flux initially
increases with the source size at a constant $\tau_{\nu}$ and then
decreases due to the decreasing optical depth as a consequence of the expansion.
For the $\sim$4$\times$10$^6$\solm ~super-massive black hole at 
the position of Sgr~A*,
one Schwarzschild radius is R$_s$=2GM/c$^2$$\sim$10$^{10}$~$m$
and the velocity of light corresponds to 
about 100~R$_s$ per hour. 
For $t > t_0$ the decaying flank 
of the curve can be shifted towards later times by first
increasing the turnover frequency $\nu_0$ or the initial source size $R_0$,
and second, by lowering the spectral index $\alpha_{synch}$ or the peak 
flux density $S_0$.
Increasing the adiabatic expansion velocity v$_{exp}$ shifts the peak of
the light curve to earlier times.
Adiabatic expansion will also result in a slower decay rate and
a longer flare timescale at lower frequencies.

\subsection{Description and properties of the SSC model}
\label{section:SSCmodel}

We have employed a simple SSC model to describe the observed 
radio to X-ray properties
of SgrA* using the nomenclature given by
Gould (1979) and Marscher (1983).
Inverse Compton scattering models provide an explanation for
both the compact NIR and X-ray emission
by up-scattering sub-mm-wavelength photons into these spectral domains.
Such models are considered
as a possibility in most of the recent modeling approaches
and may provide important insights into some fundamental
model requirements.
The models do not explain the entire low frequency radio spectrum
and the bremsstrahlung X-ray emission that dominates the IQ state.
Also high power X-ray fares (e.g. Porquet etal. 2003, 2008)
may involve additional emission mechanisms.
However, for X-ray flares of up to several 10 times the quiescent emission 
the SSC models provide a successful description of the compact IQ and 
flare emission 
originating from the immediate vicinity of the central black hole.
A more detailed explanation is also given by Eckart et al. (2004).

We assume a synchrotron source of angular extent $\theta$. 
The source size is of the order of a few Schwarzschild
radii R$_s$=2GM/c$^2$ with R$_s$$\sim$10$^{10}$~$m$ for a
$\sim$4$\times$10$^6$\solm ~black hole. One R$_s$ then corresponds
to an angular diameter of $\sim$8~$\mu$as at a distance to the Galactic
Center of 8~kpc (Reid 1993, Eisenhauer et al. 2003, Ghez et al. 2005).
The emitting source becomes optically thick at a frequency
$\nu_m$ with a flux density $S_m$, and has an optically thin spectral
index $\alpha$ following the law $S_{\nu}$$\propto$$\nu^{-\alpha}$.
This allows us to calculate the magnetic field strength $B$ and the
inverse Compton scattered flux density $S_{SSC}$ as a function of the
X-ray photon energy $E_{keV}$.  The synchrotron self-Compton spectrum
has the same spectral index as the synchrotron spectrum that is 
up-scattered 
i.e. $S_{SSC}$$\propto$$E_{keV}$$^{-\alpha}$, and is valid within the
limits $E_{min}$ and $E_{max}$ corresponding to the wavelengths
$\lambda_{max}$ and $\lambda_{min}$ (see Marscher et al. 1983 for
further details).
We find that Lorentz factors $\gamma_e$
for the emitting electrons of the order of 
typically 10$^3$ are required to produce a sufficient SSC flux in the
observed X-ray domain.
A possible relativistic bulk motion of the emitting source results 
in a Doppler
boosting factor $\delta$=$\Gamma$$^{-1}$(1-$\beta$cos$\phi$)$^{-1}$.
Here $\phi$ is the angle of the velocity vector to the line of sight,
$\beta$ the velocity v in units of the speed of light $c$, and
Lorentz factor $\Gamma$=(1-$\beta$$^2$)$^{-1/2}$ for the bulk motion.
Relativistic bulk motion 
is not a necessity to produce sufficient SSC flux density but 
we have used modest values for 
$\Gamma$=1.2-2 and $\delta$ ranging between 1.3 and 2.0 (i.e. angles $\phi$ 
between about $10^{\circ}$ and $45^{\circ}$)
since they will occur
in cases of relativistically orbiting gas as well as relativistic 
outflows - both of which are likely to be relevant to SgrA*.

\section{Modeling the light curves}
\label{section:ModelingLightCurves}

Our primary goal was to generate a model that includes 
the entire data set on the flare event observed 
on July 7, 2004 from the mm- to the X-ray domain.
Models like F1 or F2 (Eckart et al. 2006, their Table 9) 
or the dynamical, multicomponent model presented by
Eckart et al. (2008a) reproduce the NIR/X-Ray 
properties of the observed flare $\phi$3/III and
$\phi$4/IV very well.
We have repeated this modeling under the premise of achieving
fits with a simultaneous match to the SMA and VLA data.

The six so far reported coordinated SgrA* measurements
that include sub-mm data 
(Eckart et al. 2006b; Yusef-Zadeh et al. 2006b; Marrone et al. 2008, 
Eckart et al. 2008b) have
shown that the observed submillimeter flares follow strong NIR or X-ray events.
If the events were unrelated we would expect an equal number of
submillimeter flares leading and following the NIR/X-ray events
(see detailed discussion in Marrone et al. 2008).
We therefore assume that the sub-millimeter 
flare presented here is related to the observed IR flare events.

For the July 7 SMA and VLA data Eckart et al. (2006a) have shown 
that the observed amplitudes of the flux density variations are 
generally consistent with adiabatic expansion of a 
synchrotron self-absorbed source (van der Laan 1966).
Following the THz peaked NIR/X-ray flare events 
III/$\phi3$ and IV/$\phi4$ on July 7 (see Fig.~\ref{Fig:NIRXraydata} and 
modeling results given by Eckart et al. 2006a)
the radio flux density will first rise and later drop
as the source evolves.

However, a detailed comparison to theoretical light curves of 
adiabatically expanding source components shows that
a single source component cannot give a satisfactory fit to the
sub-mm/mm data.
In Fig.~\ref{Fig:single} we show that when the 340~GHz light curve is 
decaying, the 43~GHz curve is still rising (straight bold face 
section in the correponding lines).
When the 43~GHz curve is decaying then the 
decaying 340~GHz light curve is at flux density levels
well below the 43~GHz curve (dashed bold face 
section in the correponding lines).
Both scenarios are inconsistent with the observations on July 04.
This result is also independent of the expansion velocity.
Modeling the 2004 July 07 radio data therefore must involve a 
minimum of two source components.
Quite naturally two components can be associated with the 
NIR/X-ray flares  III/$\phi$3 and IV/$\phi$4.

Our goal is to fit the variable part of the sub-mm/mm light curves
(see section \ref{section:SMA}) with a source model that is 
also able to describe the observed NIR/X-ray properties.
We calculated model light curves at 340~GHz and 43~GHz for the model 
with a smallest (4) and larger (6) number of source components.
Smaller numbers of components cannot account for all essential features of 
the sub-mm/NIT/X-ray light curves. A detailed explanation is given in
the following.

\begin{figure}
\centering
\includegraphics[width=8.5cm,angle=-00]{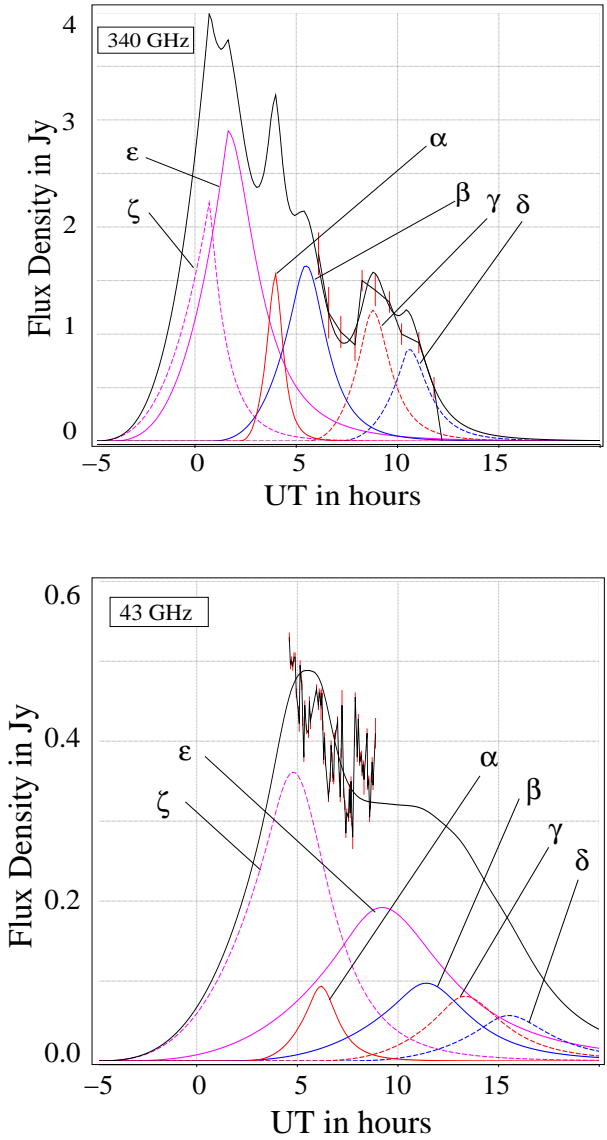}
\caption{\small Decomposition of the 340~GHz and 43~GHz light curve for model
B1 (see Table~\ref{modeldata}) into the contributions 
of the different source components ($\alpha$ to $\zeta$).
The data are shown with the offsets discussed in
sections \ref{section:SMA} and \ref{section:VLA} removed.
}
\label{Fig:compose}
\end{figure}

\begin{figure}
\centering
\includegraphics[width=8.5cm,angle=-00]{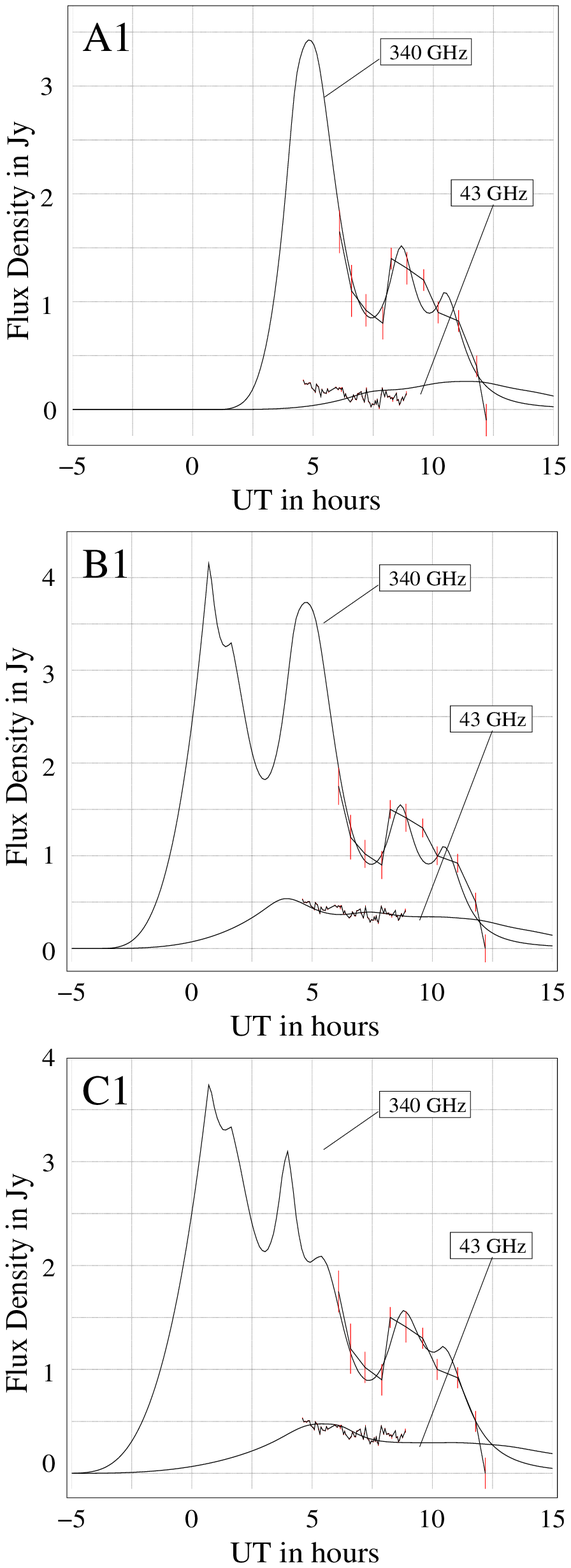}
\caption{\small  The observed 43~GHz and 340~GHz light curves 
shown in comparison to model A1, B1, and C1.
The data is shown with the offsets discussed in
sections \ref{section:SMA} and \ref{section:VLA} removed.
}
\label{Fig:model1}
\end{figure}

\begin{figure}
\centering
\includegraphics[width=8.5cm,angle=-00]{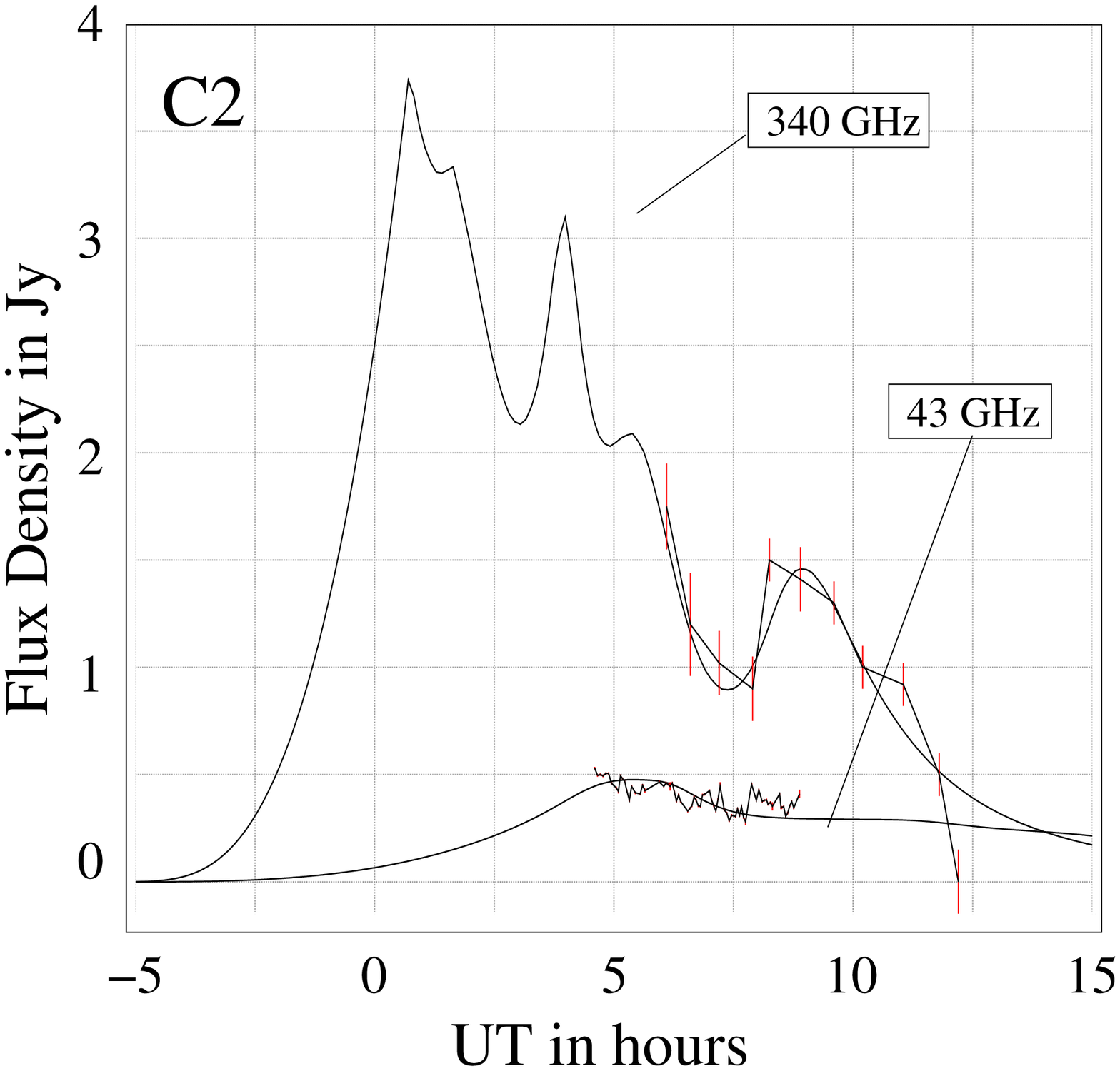}
\caption{\small  The observed 43~GHz and 340~GHz light curves 
shown in comparison to model C2.
The data is shown with the offsets discussed in
sections \ref{section:SMA} and \ref{section:VLA} removed.
}
\label{Fig:model2}
\end{figure}

In Fig.~\ref{Fig:compose} we show the decomposition of the overall light curves
into the contribution of individual source components.
In Figs.~\ref{Fig:model1} and \ref{Fig:model2} we show the model light curves 
for comparison to the measured 340~GHz and 43~GHz data.

{\it SSC modeling with adiabatically expanding source components:}
We iterated between the SSC modeling of the NIR/X-ray data and the
modeling of the sub-mm/mm data as adiabatically expanding sources using the
same component parameters as given in Table~\ref{modeldata}.
The source expansion is also motivated by an indication of 
a hot spot evolution within 
a possible accretion disk based on a May 2007 flare event.
Eckart et al. (2008a) have described the July 2004 flare using a 
multi-component  disk model allowing for a source size increase of
(at least) 30\% over about 40 minutes in order to explain the strong
decrease of the X-ray flux density between $\phi$3 and $\phi$4.

In Table~\ref{modeldata} we summarize the properties of 6 different models
that we considered to represent the 2004 July 7 radio, NIR,and X-ray data.
For each model the 'x' symbols in column 2 and 3 indicate which of the source components
have been considered for the models. Correspondingly they are labeled 
A1, A2, B1, B2, C1 and C2.
The 340~GHz flare SMA5 is accounted for by either 
a double ($\gamma, \delta$) or a single ($\eta$) component 
in A1, B1, C1 and A2, B2, C2, respectively, 
using the source labels marked in columns 1, 2 and 3.
The decaying 43~GHz flux density component VLA1
is accounted for by simple flux density offsets in models A1 and A2 (see 
caption of Table~\ref{modeldata}) or by 2 components ($\epsilon, \zeta$)
in models (B1, B2, C1, C2).
In column 4 the individual adiabatically expanding source components are labeled
and identified with the flares detected in the different wavelength regimes
using the nomenclature by Eckart et al. (2006a).
The modeling was done with constant expansion velocities of 0.005~c for
models (A, B) and 0.006~c for models (C). Results of the modeling are 
discussed in section~\ref{section:Modelingresults}.

The predictions from the SSC-modeling and of
the optically thin NIR flux density from the sub-mm data are 
especially sensitive
to variations of the model parameters.
The uncertainties of the model parameters given in the first row of
Table~\ref{modeldata} were derived from
a comparison of observed and predicted NIR and X-ray model flux densities
and from
reduced $\chi^2$ values calculated by comparing the
SMA and VLA data with the adiabatic expansion models.

A global variation of a single parameter 
by the value listed in the corresponding column
results in an increase of $\Delta \chi = 1$.
Here global variation  means:
adding for a single model parameter but for all source components the 1$\sigma$
uncertainty, such that a maximum positive or negative flux density deviation is reached.

Alternatively, a variation by the listed uncertainty for 
only a single source component results in a variation of the model
predicted NIR and X-ray flux density by more than 30\%.
Judging from the $\Delta \chi$ based on the sub-mm data only
the global uncertainties for $S_{max,obs}$, $\alpha_{synch}$ and $R_0$ 
could be doubled.

The minimum number of source components is 5.
In the reduced $\chi^2$ fit we used 5 times 4 ($S_{max}$, $\alpha_{synch}$,
$R_0$, $\nu_{max}$) plus one common expansion velocity $v_{exp}$ and
time offset (leaving the time differences between the components fixed),
i.e. 22 degrees of freedom.
The model parameters can unfortunately not all be considered as being
independent, e.g. the width and peak of a light curve signature
depends to a varying extend on all 4 parameters $S_{max}$, $\alpha_{synch}$,
$R_0$, and $\nu_{max}$. Therefore we stayed for all models
with the minimum number of source components (5) to estimate
the degrees of freedom. Leaving all parameters free ($\Delta$t, $v_{exp}$,
$S_{max}$, $\alpha_{synch}$,
$R_0$, and $\nu_{max}$) would double the number of degrees of freedom
and reduce the $\chi^2$ values correspondingly.
Since the VLA data consists of 5 to 6 times the number of SMA data points
we weighted the squared SMA flux deviations and number of data points
by factor of 6. The $\chi^2$ test was then carried out using the
sum of the squared flux deviations and data points of the VLA and SMA
datasets.

Of course the models have been set up under the constraint of minimizing the
number of free parameters (and to maximize the description of significant 
flare features in the observed light curves).
The flux excursions labeled (VLA1,$\phi$2/3) could easily be explained 
by 3 or 4 components rather than 2 ($\epsilon$ and $\zeta$) in the same way
as SMA5 can be explained by 2 ($\gamma$ and $\delta$) rather than a single 
component ($\eta$).

In the following we comment on the detailed modeling of different sections 
of the light curves.

{\it Modeling of individual portions of the light curves:}

Significant NIR and X-ray flux density is only produced from the initial
THz peaked source components $\alpha$ and $\beta$. 
Using a linear expansion at constant speed v$_{exp}$ we then calculate 
light curves in the selected sub-mm/mm bands.
For an expansion speed of $\sim$0.005~c we find that the radio model component
$\beta$ (due to the III/$\phi3$ NIR/X-ray flare event) can account for 
a major portion of the decreasing part of the 340~GHz light curve 
- SMA4 in Fig.~\ref{Fig:SMAVLAdata} -
between 06 and 12 UT on July 2004.

Components $\alpha$ and $\beta$ also account for 
a portion of the final section 
(after about 6 hours UT - VLA1 in Fig.~\ref{Fig:SMAVLAdata})
of the decreasing 43~GHz light curve.
For models A and B the spectral indices of components $\alpha$ and $\beta$ 
of  $\alpha_{synch}$=0.8 are consistent with the value of
0.6$\pm$0.2 found by Hornstein et al. (2007; see discussion in 
section~\ref{section:NIRfluxes}).
While the NIR flux density in models (A, B) is provided by direct synchrotron emission,
the flux density in model (C) is produced via the low frequency section of the
scattered SSC specrum. In this case the spectral index must be steep 
($\alpha_{synch}$=1.2) and is close to the 
value derived for the overall NIR/X-ray spectral index of $\alpha$ and $\beta$
of $\alpha_{NIR/X-ray}$=1.2$\pm$0.2  (Eckart et al. 2006a).
This is consistent with the fact that the optically thin synchrotron
spectral index (sub-mm to NIR) is expected to equal the broad band spectral
index of the SSC spectrum.

Components $\alpha$ and $\beta$ cannot fully account for the 43~GHz radio flux density.
This is especially true for the the initial section (before 6 hours UT) of 
the decreasing 43~GHz light curve.
The remaining flux density of the VLA1 flare event then has to be explained by
source components that provide an almost flat light curve between 5 and 9 hours UT.
Alternatively, source components that can be identified with the 
the NIR flare component II/$\phi$2/3 (see Fig.~\ref{Fig:NIRXraydata})
are required to deliver this flux density contribution. 
To model the light curve in this region a minimum of 2 model components
($\zeta$ and $\epsilon$) is needed.
The observed flux densities (or limits) and implied time differences
require rather steep spectral indices 
of $\alpha_{synch}$=1.3 for these source components. 
Especially for models C1 and C2 the predicted peak flux densities for the 
340~GHz and 43~GHz bands lie well within 
the range of observed flux density values 
(e.g. Marrone et al. 2008, Yusef-Zadeh et al. 2008, Eckart et al. 2006a).

The 340~GHz flare event SMA5 requires at least one ($\eta$) or 
two model components ($\gamma$, $\delta$).
For two source components a steep
spectral index with $\alpha_{synch}$=0.9-1.4 results in a 
weak NIR flare event with an X-ray flux that would have been non-detectable 
for Chandra in the presence of the quiescent X-ray bremsstrahlung component of 
SgrA*.
For a single component a flat spectral index of $\alpha_{synch}$=0.2
(corresponding to $p$=0.6) is required. This results in a 
significant NIR/X-ray flare about 4$\pm$1 hours after the III/$\phi3$ flare event.
Fig.6 in Eckart et al. (2006a) shows that no X-ray event was detected 
by Chandra over this period of time.
Therefore a flat spectral component $\eta$ can be excluded.

\begin{figure}
\centering
\includegraphics[width=8.5cm,angle=-00]{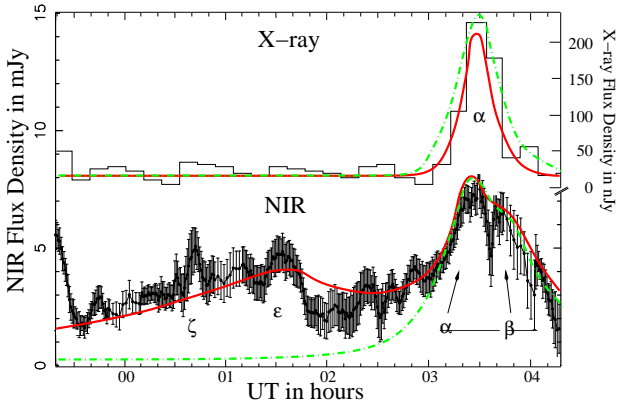}
\caption{\small  
A comparison between the X-ray and NIR 2.2 $\mu$m light curves 
shown in Fig.~\ref{Fig:NIRXraydata} and the modeling results for the
A, B (both in green), and C (red) models. 
The contributions of individual model components 
as listed in Table~\ref{modeldata} are labeled.
}
\label{Fig:fitNIRXraydata}
\end{figure}

\subsection{Discussion of the modeling results}
\label{section:Modelingresults}

Our modeling shows that the observed mm through X-ray data of the
July 7, 2004 flare event can be successfully modeled by a combination of
an SSC and an adiabatic expansion model.
While our analysis cannot fully support or even prove the model of
adiabatic expansion for the flare emission of SgrA* by itself, it is in 
full agreement with model results that have been obtained by analyzing 
previous observations
(e.g. Eckart et al. 2006a, 2008b, Yusef-Zadeh et al. 2007, 2008, 
Marrone et al. 2008).
The modeling also suggests the presence of two flare phases in the 
simultaneous NIR/X-ray events.
The fact that a single component adiabatic expansion model cannot
account for the observed sub-mm/mm light curves, 
the considerable spread in source parameters 
(brightness, size, spectal index, sub-mm turnover frequency)
as well as the presence of a non rapidly variable 
(on timescale of more than one or a few days) flux density component
indicates low level activity in addition to the presence of repeated
individual bright flares.
The flare event II/$\phi$2/3 probably is an example of such 
low level activity.

\begin{figure}
\centering
\includegraphics[width=8.5cm,angle=-00]{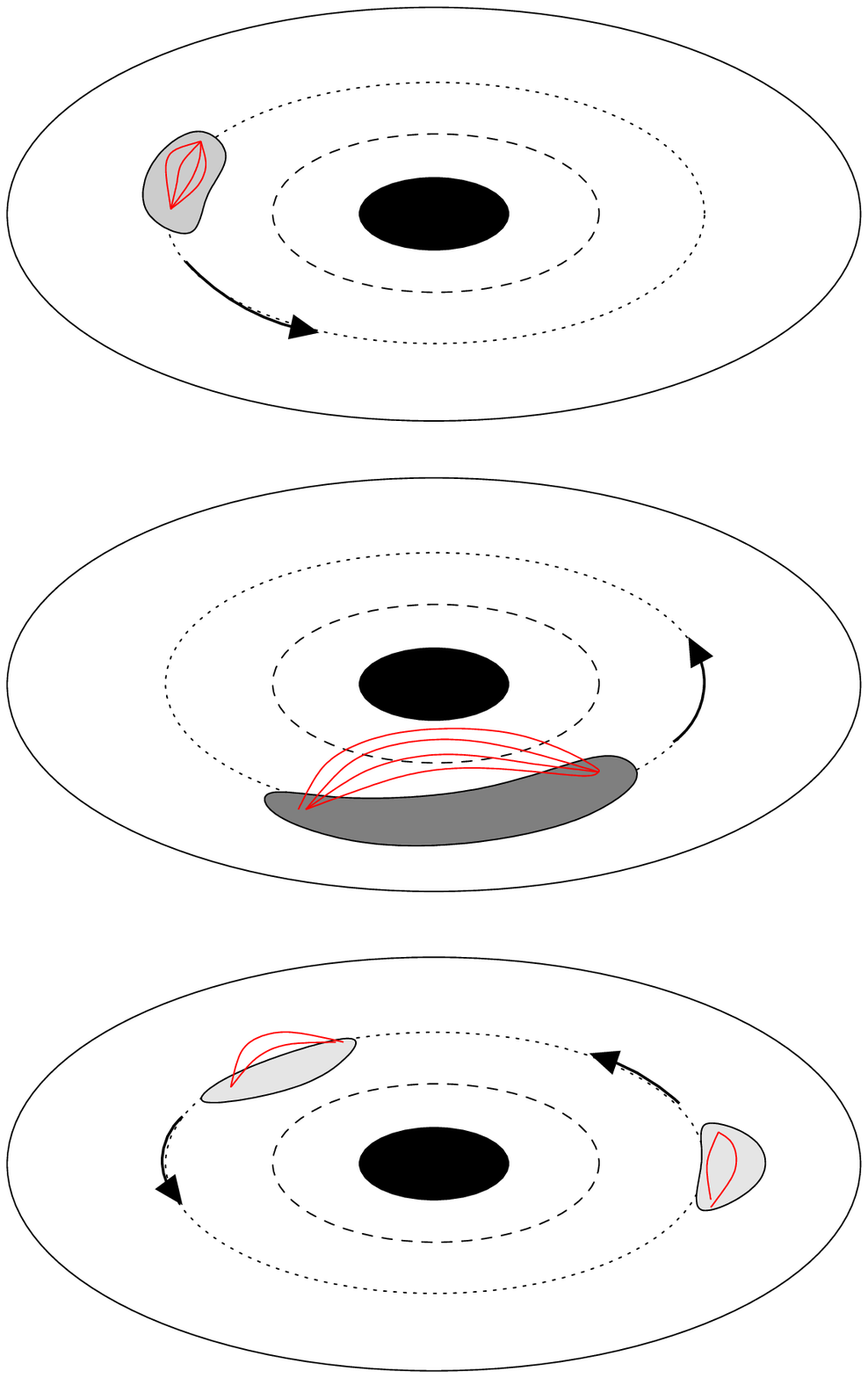}
\caption{\small  
Sketch of an expanding hot spot within an inclined
temporary accretion disk of SgrA* based on 
Hawley \& Balbus (1991, see also Balbus \& Hawley 1998 and Balbus 2003).
The black center indicates the event horizon of the massive black hole, the solid line
the outer edge of the accretion disk (see e.g. Meyer et al. 2005),
The long dashed line marks the inner last stable orbit. The dotted line represents
a random reference orbit to show the effect of differential rotation of an extended emission region.
The red solid line across the grey shaded extended spots depicts the magnetic field
lines that through magneto-hydrodynamical instabilities
provide a coupling between disk sections at different radii.
}
\label{Fig:diskmodel2}
\end{figure}

Our modeling also requires the low adiabatic expansion velocity  
that describes other flares of SgrA*
(Yusef-Zadeh et al. 2007, 2008, Eckart et al. 2008b).  
Our values of $\sim$0.005~c fits well into the range of velocities 
of $v_{exp}$=0.003-0.1c given by Yusef-Zadeh et al. (2008).  
It is currently unclear how to interpret the expansion velocity.
The velocities are lower than the orbital velocities for 
flare components introduced from orbiting spot models and accretion disk
models 
(Eckart et al. 2006b, 2008ab, Meyer et al. 2006ab, 2007, 2008, Trippe et al. 2007).
The observed sub-mm/mm flare emission therefore can be 
explained as an expansion within the occasionaly present accretion disk around SgrA*
(as shown in Fig.~\ref{Fig:diskmodel2}) or a flaring of the disk corona.

We can now derive the delay times to the sub-mm/mm bands 
relative to the NIR/X-ray flare times from our modeling.
A summary is given in Table~\ref{delaydata}.
The delay times of the compact ($\alpha, \beta$) 
components with cutoff frequencies in the THz domain are 
1-2 hours in the 340~GHz and  3-9 hours in the 43~GHz band. 
The delay  times of the more extended source components 
($\gamma$ to $\eta$)
with cutoff frequencies at several 100~GHz are
about 20 minutes in 340~GHz and  4-7 hours in 43~GHz band. 
In general larger source components ($>$1.5~R$_s$) 
with low peak flux densities (S$_{max,obs}$$<$3~Jy) 
and low cutoff frequencies ($\nu_{max,obs}$$<$a few 100~GHz) 
show little or no X-ray emission and have 
shorter delay times towards the 345~GHz band (see also Marrone et al. 2008).

In Fig.~\ref{Fig:diskmodel2} we show how a double or extended spot model can be 
interpreted in an evolutionary framework for disk structure.
This model is based on work by 
Hawley \& Balbus (1991, see also Balbus \& Hawley 1998 and Balbus 2003)
and is motivated by the fact that accretion
disks show magneto-rotational instabilities in which
magnetic field lines provide a coupling between disk sections at different
radii resulting in an efficient outward transport of angular momentum.
In a differentially rotating disk the inner disk portions that lose 
angular momentum 
will slide into lower lying orbits, and rotate more rapidly.
The center image in Fig.~\ref{Fig:diskmodel2} represents the 
expanded disk component that may be the source of a flaring corona or 
inhanced jet activity since it is located at the footpoint of a possible short
jet that may be associated with SgrA*.
The occasional presence of a thin disk in which the Hawley \& Balbus
mechanism is operating effectively may imply an accretion rate and 
corresponding luminosity higher than that expected for an accretion flow 
that is otherwise assumed to be radiatively inefficient.
This may support that the actual accretion rate is indeed a strong function 
of the radius and much below the rate expected from the mass loss of the
surrounding He-stars (see references in introduction).

Alternatively, the expanding source components may also have a
bulk velocity that can be substantially larger than the
observed expansion velocity.
As the jet or wind expands, different regions of it will dominate 
the emission at successively lower frequencies (see 
Falcke \& Markoff 2000, 2001, Yuan et al. 2002, 
as well as Fig.12 by Eckart et al. 2008a).

\subsubsection{Matching the NIR flux densities}
\label{section:NIRfluxes}

We have re-modeled the July 7, 2004 flare to obtain a 
smaller discrepancy between the measured and predicted NIR flux densities.
Our models A and B for the light curve features
$\phi$3/III and  $\phi$4/IV overestimate the 
NIR flare flux by a factor of 4 (Table~\ref{modeldata}) to 7 (Eckart et al. 2006a).
Since direct synchrotron emission from $\gamma_e$$\sim$10$^3$ electrons 
delivers the dominant portion of the NIR flux density 
this discrepancy becomes larger if we use flatter spectral indices 
like $\alpha$$\sim$0.4 (Model F2 by Eckart et al. 2006a). 
This effect cannot be compensated for 
using a lower turnover frequency $\nu_m$ or flux density $S_m$ since then the
X-ray emission cannot be matched any longer.

The NIR flare emission shows observed spectral indices of
$\alpha$$\sim$0.6 (Ghez et al. 2005, Hornstein et al. 2006) or even
steeper (Eisenhauer et al. 2005, Gillessen et al. 2006, Krabbe et al. 2007).
Eckart et al. (2006a) assume that steep NIR spectral indices may be a 
consequence of synchrotron losses represented by an exponential cutoff
in the energy spectrum of the relativistic electrons (see also Liu et al. 2006).
This will appear as a modulation of 
the intrinsically flat spectra 
with an exponential cutoff proportional to $exp[-(\lambda_0/\lambda)^{0.5}]$ 
(see e.g. Bregman 1985, and Bogdan \& Schlickeiser 1985)
and a cutoff wavelength $\lambda_0$ in the infrared.
If $\lambda_0$ lies in the 4-8$\mu$m wavelength range, then the variation in the spectral 
index is of the order of $\Delta$$\lambda$=0.6-1.0 and can easily explain 
the factors of 4 to 10 between the observed and modeled NIR flux density.
This is the case for models A and B.
Small variations in such an exponential damping of the radiation provide
variable infrared  spectra with spectral indices that may be 
consistent with the observed values.
Such a scenario explains NIR flux densities that fall below the 
values predicted by the SSC model.
In model C the flux density is produced via the low frequency 
section of the scattered SSC spectrum  with a steep 
spectral index  ($\alpha_{synch}$=1.2).
This solution implies no NIR polarization of source component $\alpha$
but provides a satisfactory fit to the NIR and sub-mm/mm 
flux densities.

In comparison to the data
plotted in Fig.~\ref{Fig:NIRXraydata} we show in
Fig.~\ref{Fig:fitNIRXraydata} the NIR- and X-ray light curves
that correspond to the models listed in Table~\ref{modeldata}.
The SSC-modeling was done referring solely to the
peak values of the individual flare features in the measured light curves.
While in models A and B the components $\epsilon$ and $\zeta$ have not been
considered (to minimize the number of free parameters) or
result in very small contributions to the NIR and X-ray flux densities,
model C reproduces the overall shape of the light curves quite
satisfactorily. We have not plotted the NIR- and X-ray contributions of
components $\gamma$, $\delta$, or $\eta$ that contribute to the
sub-mm flare SMA5, since there are no NIR observations available
(see comment on SMA5 at the end of the opening of section
\ref{section:ModelingLightCurves}).
Therefore, in Fig.~\ref{Fig:NIRXraydata} it is also not necessary
to distinguish between models with different model labels (1 or 2)
as listed in Table~\ref{modeldata}.

\begin{table}
\begin{center}
\begin{tabular}{lccccccc}\hline 
component & NIR/      &340~GHz & 43~GHz  & delay& delay   \\
          & X-ray     &    &     &  340~GHz &  43~GHz   \\ \hline
$\alpha$  & 3.3       & 4  & 6   &  0.7     &2.7   \\
$\beta$   & 3.8       &5.5 &11.5 &  1.7     &7.7   \\
$\gamma$  & 8.3       & 9  &13   &  0.7     &4.7   \\
$\delta$  & 10.4      & 11 &16   &  0.6     &5.6   \\
$\epsilon$& 1.6       & 2  & 9   &  0.4     &7.4   \\
$\zeta$   & 0.7       & 1  & 5   &  0.3     &4.3   \\
$\eta$    & $\sim$9.0 & 10 &15   &  1.0     &6.0   \\
\hline 
\end{tabular}
\caption{
Columns 2 to 4 list the UT times in hours of the NIR/X-ray, 
340~GHz and 43~GHz flux desity peaks for source components
of models A,B, and C with properties listed in Table~\ref{modeldata}
Columns 5 and 6 list the delay times in hours from
the time of birth to the 340~GHz and 43~GHz  frequency bands.
The time of origin of the components at their peak frequency 
are synchronous with the NIR/X-ray flare times.
The uncertainties are of the order of one hour.
\label{delaydata}}
\end{center}
\end{table}

\begin{table*}
\centering
{\begin{small}
\begin{tabular}{lllclrrrrrrrrrr}
\hline
model & 1 & 2 &  source   & flare & $\Delta$t &v$_{exp}$ &S$_{max, obs}$&$\alpha_{synch}$&R$_0$&$\nu_{max, obs}$ & B &S$_{NIR, synch}$ &S$_{NIR, SSC}$  &S$_{X-ray, SSC}$\\
lable &   &   &           & lable &    hours & in $c$ &[Jy]            &        &     & [GHz]       & [G]          & [mJy]             &[mJy]             &[nJy]             \\
 \hline
      &   &   &           &                &     &      &     &     &     &      &  &     &        &        \\
$1~\sigma$ $\rightarrow$  &   &   &  &  & $\pm$1.0 &$\pm$0.001 &$\pm$0.1&$\pm$0.1 &$\pm$0.1 &$\pm$250 &$\pm$10&$\pm$1.0&$\pm$1.0 &$\pm$20 \\
      &   &   &           &                &     &      &     &     &     &      &  &     &        &        \\
 \hline
      &   &   &           &                &     &      &     &     &     &      &  &     &        &        \\
A~~~B    & x & x & $\alpha$  & $\phi$3        &  0.0& 0.006& 9.0 & 0.8 & 0.9 & 1750 &50& (39)& $<$1.0 & 230    \\
A~~~B     & x & x & $\beta $  & SMA4, $\phi$4  & +0.5& 0.006&11.2 & 0.8 & 1.8 & 1230 &63& (34)& $<$1.0 & $<$10  \\
A~~~B     & x & - & $\gamma$  & SMA5           & +5.0& 0.006& 2.4 & 0.9 & 1.5 &  680 &65& 4.5 & $<$1.0 & $<$10  \\
A~~~B     & x & - & $\delta$  & SMA5           & +7.1& 0.006& 1.2 & 0.9 & 1.8 &  525 &63& 1.5 & $<$1.0 & $<$10  \\
~~~~~~B     & x & x & $\epsilon$& VLA1, $\phi$2/3& -1.7& 0.006& 3.0 & 0.9 & 2.9 &  470 &18& 0.0 & $<$1.0 & $<$10  \\
~~~~~~B     & x & x & $\zeta$   & VLA1, $\phi$2/3& -2.6& 0.006& 3.9 & 1.4 & 2.9 &  223 & 1& 0.0 & $<$1.0 & $<$10  \\
A~~~B     & - & x & $\eta  $  & SMA5           & +4.9& 0.006& 4.0 & 0.2 & 1.2 & 1350 &85& 600 & $<$1.0 & 390    \\
      &   &   &           &                &     &      &     &     &     &      &  &     &        &        \\
 \hline 
      &   &   &           &                &     &      &     &     &     &      &  &     &        &        \\
{\bf C}&{\bf x}& x & $\alpha$  & $\phi$3        &  0.0& 0.005 & 8.2 & 1.2 & 0.6 & 1310 &23& 0.0 & 7.8    & 190    \\
       &{\bf x}& x & $\beta $  & SMA4, $\phi$4  & +0.5& 0.005 & 8.5 & 1.2 & 1.6 & 1310 &68& 5.5 & 0.5    & $<$10  \\
       &{\bf x}& - & $\gamma$  & SMA5           & +5.0& 0.005 & 2.4 & 0.9 & 1.5 &  680 &50& 5.1 & $<$1.0 & $<$10  \\
       &{\bf x}& - & $\delta$  & SMA5           & +7.1& 0.005 & 1.2 & 0.9 & 1.8 &  525 &57& 1.6 & $<$1.0 & $<$10  \\
       &{\bf x}& x & $\epsilon$& VLA1, $\phi$2/3& -1.7& 0.005 & 3.5 & 0.9 & 2.9 &  380 &16& 4.0 & $<$1.0 & $<$10  \\
       &{\bf x}& x & $\zeta$   & VLA1, $\phi$2/3& -2.6& 0.005 & 3.0 & 1.4 & 2.9 &  220 & 4& 0.1 & $<$1.0 & $<$10  \\
       &   -   & x & $\eta  $  & SMA5           & +3.2& 0.005 & 2.9 & 0.2 & 1.5 & 1060 &55& 390 & $<$1.0 & 270    \\
       &       &   &           &                &     &      &     &     &     &      &  &     &        &        \\
 \hline 
\end{tabular}
\end{small}}
\caption{
Source component parameters for the combined SSC and adiabatic expansion model
of the 7 July, 2004 flare. 
The offset times $\Delta$t are given with respect to the peak of the brighter NIR flares $\phi$3
(synchronous with the brightest X-ray flare III) at about 7 July, 204, 03:15:00 UT.
The adiabatic expansion of the individual source components occurs at a constant velocity
of 0.006~c for model A and B and of 0.005~c for model C.
In column 4 we identify the observed flares feartures that are represented by the corresponding
model component, using the nomenclature introduced in Eckart et al. (2006a) and here (see Fig.~\ref{Fig:NIRXraydata}). 
The different source component models A, B, C discussed in the text are labeled in column 1.
The flare components required for models A1, B1, C1 and A2, B2, C2 are marked with an 'x' in column 2 and 3, respectively.
Flux density offsets of the VLA and SMA data used for models A1/A2 are 0.26~Jy and 1.9~Jy.
Flux density offsets of the VLA and SMA data used for models B1/B2/C1/C2 and  are 0.0~Jy and 1.8~Jy,  respectively.
The $\Delta \chi$ values in comparison to the mm/sub-mm data that we obtained for the various models are:
A1: 2.9; 
A2: 3.1; 
B1: 1.71; 
B2: 1.81; 
C1: 1.70; 
C2: 1.80.
\label{modeldata}}
\end{table*}

\subsubsection{Matching the sub-mm/mm flux density offsets}
\label{section:submmoffsets}

Models A1 and A2 fail to fit the VLA data at 43~GHz.
In particular the slope and the first part of the data are not matched.
Higher expansion velocities help to match the 43~GHz data but fail to
match the first decaying flank of the 340~GHz SMA data.
In addition it is required that a considerable part of the
variable 43~GHz flux density has to be modeled as a 
constant offset in addition to the 1.4~Jy as the lowest 
visibility flux density (see section~\ref{section:VLA}).
Therefore, also given the poor match of the NIR flux densities of light curve 
components $\alpha$and $\beta$,
models A1 and A2 - although requiring the smallest number of 
source components - are not very likely as an explanation of the
observed mm/sub-mm light curves.

With S(43~GHz)$\sim$1.7~Jy and S(340~GHz)$\sim$2.4~Jy  
the spectral index is $\alpha_{43/340} \sim -0.17$ indicative
of an inverted spectrum that must be related to
a compact source component. This is inconsistent with the assumption
that these flux density contributions are due to a
non-  or only slowly variable source component. 
Therefore, non-expanding synchrotron source components
are not a satisfying explanation for the constant flux density 
contributions.

We may, however, assume that the fluxes of the non-variable part 
of the light curve are due to a more extended component.
If that component is also self-absorbed and adiabatically expanding
then the peak flux densities will be
$S(\nu_2)=S(\nu_1)[\nu_2/\nu_1]^{(7p+3)/(4p+6)}$~
(van der Laan 1966).
With S(43~GHz)$\sim$1.4~Jy and S(340~GHz)$\le$2.4~Jy we find 
$\alpha$$\sim$0.26 or p=1+2$\alpha$=1.52 which is between the 
value of p=2.2 expected for a spectral index of 0.6
and a value of p=-0.1 found 
prior to the flare peak on July 17, 2006, reported by 
Marrone et al. (2008).

We conclude that even if we use the smallest number of source components
to describe the 2004 flare emission the residual flux densities 
that are not taken into account by the model at both frequencies 
are consistent with adiabatic expansion of source components.

\subsubsection{The magnetic field}
\label{section:magneticfield}

The magnetic field strengths between
5 and 70 Gauss (Eckart et al. 2006a, 2008a, Yusef-Zadeh et al. 2008) are consistent with
sub-mm/mm  variability timescales of synchrotron components with
THz peaked spectra and the assumption that these source components 
have an upper frequency cutoff $\nu_2$ in the NIR, i.e. that they contribute significantly to
the observed NIR flare flux density.
Here the upper frequency cutoff to the synchrotron spectrum is assumed to be at 
$\nu_2 = 2.8 \times 10^6 B \gamma_2$ in Hertz, with the magnetic field strength in Gauss.
The Lorentz factor $\gamma_2$ corresponds to the energy $\gamma_2 mc^2$ at the upper
edge of the electron power spectrum.
For $\gamma_2$$\sim$10$^3$ and B around $60~G$, the synchrotron cutoff falls into the NIR.

A comparison between typical flare timescales and synchrotron cooling
timescales can also be used to derive estimates of the required magnetic
field strengths. 
The flux density variations of Sgr~A* can be explained 
in a disk or jet model (see e.g. discussion in Eckart et al. 2006ab, 2008a), or 
they could be seen as a consequence of an underlying physical process that can 
mathematically be described as red-noise
(Do et al. 2008, Meyer et al. 2008).

In the orbiting spot model this timescale will reflect the flux 
modulation by the relativistic orbital motion of the spots.
While the overall flare length is of the order of 2~hours
(Eckart et al. 2006a)
shorter timescales of about 20 minutes can be attributed to the sub-flares 
(Genzel et al. 2003, Eckart et al. 2006b).
However, the spot lifetime is likely of the order of the orbital
timescale or even shorter
(Schnittman 2005, Schnittman et al. 2006, Eckart et al. 2008a).
In order to match the overall typical flare timescale of about
2 hours and given a minimum turnover frequency around 300~GHz 
the minimum required magnetic field strength 
is of the order of 5 Gauss.
This is required as a minimum value to have the cooling time of 
the overall flare less than the 
duration of the flare (Yuan, Quataert, Narayan 2003, 2004, Quataert 2003).
Similarly, as the flare expands it will cool on the synchrotron 
cooling timescale
$t_s \sim 3 \times 10^7 \nu_9^{-0.5} B^{-3/2}$,
where $t_s$ is in seconds, B is in Gauss, $\nu_9$ is frequency in GHz
(Blandford \& K\"onigl 1979).
Here B is the magnetic field of the synchrotron component with 
turnover frequency $\nu_m$,
turnover flux density $S_m$ and angular source size $\theta$.
Adopting a typical NIR flare timescale of 20~minutes and $\nu_9$$\sim$1000 we get B$\sim$80~G.
For much shorter spot timescales as theoretically indicated 
(Schnittman 2005, Schnittman et al. 2006) the field strengths may be even higher.
As we show in the present paper, the description of NIR/X-ray flares as SSC source 
components with THz peaked spectra
is also compatible with the assumption of adiabatic expansion of the same 
component that then can explain the observed sub-mm/mm flare emission.

The assumption of variable synchrotron components that
become optically thick in the MIR or even NIR is problematic.
For the magnetic field we find $B \sim \theta^4 \nu_m^5 S_m^{-2}$.
Given the higher NIR turnover frequency ($\sim$150~THz rather than 1~THz) 
and lower flux densities (mJy rather than Jy) 
requires source sizes much less than a Schwarzschild radius 
in order to allow for magnetic field strengths of the order of 60~G or even below.
In addition we find that with the low values for $S_m$ the adiabatic expansion 
of this source component cannot explain the sub-mm/mm flux densities. 

For the outer, larger components $\epsilon$ and $\zeta$ which are also much less constrained by
the NIR and X-ray data, we find values for the magnetic field well below 60~G.
For models (A,B) this is - to first order - consistent with an expected 
decay of the field strength in an adiabatic 
expansion model (van der Laan 1966) proportional to $R^{-2}$.
These components also require steeper spectral indices as expected - if there is a 
correlation between NIR flare strength and spectral index 
(Ghez et al. 2005,
Gillessen et al. 2006,
Krabbe et al. 2006,
Hornstein et al. 2006,
but see Hornstein et al. 2007)
as is also discussed in Eckart et al. (2006a) and  Bittner et al. (2007).
Larger sizes and steeper spectra for these source components are required 
in models B1 and B2 to
match the early section of the 43~GHz VLA data 
and to fulfill
the flux density measures and limits in the NIR and X-ray domain.

\subsubsection{Bulk motion and detectability of structure}
\label{section:bulkmotion}

VLBI experiments at mm-wavelengths have revealed a size limit for SgrA* of about 
0.5~AU corresponding to $\sim$5 ~$R_S$ (Doeleman et al. 2008, see also 
Doeleman et al. 2001, Shen 2006, Huang et al. 2007).
If the expansion speed of $\sim$0.01~c is taken as
a bulk motion then the time to cross 0.5~AU is less than 7 hours which is well above
the flare timescale. The flare flux due to the moving source component will then 
not lead to a detectable structure, extended on scales that can currently 
be probed with VLBI techniques.
Assuming a flare timescale of 2 hours this also results in an upper limit of
the bulk motion of 0.07~c to lead to a detectable structure if the VLBI size 
measurements are taken during a flare at mm-wavelengths.
At this speed, however, only plasma at distances of $\ge$11~R$_s$ can leave 
the $\sim$4$\times$10$^6$\solm black hole.
Eventually, the relativistic electrons will contribute to a possible 
overall jet or outflow with a low surface brightness.
Falcke \& Markoff (2000, 2001) and Yuan et al. (2002)
propose that the emission from Sgr~A* arises primarily in a jet. 
In this picture, a small fraction of the accretion flow is ejected near the black hole 
as a short, luminous jet.
A source structure in which an accretion disk is associated with a short jet
may explain most of the observed properties of SgrA*. 
Jet structures are associated with almost all galactic nuclei.
For the case of SgrA* such a configuration is sketched 
in Fig.9 of Eckart et al. (2008a).

It is therefore possible that the emergent spectrum of Sgr~A* is the sum of
the emission from a jet and an underlying accretion process. 
With increasing distance from Sgr~A* the plasma in
the jet becomes optically thin at ever longer wavelengths. Hence,
radiation at different radio wavelengths probes different sections of
the jet and results in a correlation between the emission
at different wavelengths.
Emission at sub-millimeter wavelengths arises at the smallest
scales, at the foot of the jet at distances of a few
Schwarzschild radii from the black hole. 
In the immediate vicinity of the black hole it is hard to
distinguish between emission from an accretion flow and from the foot
of a jet.
Details of expected low surface brightness jet geometries are discussed by 
Markoff, Bower \&  Falcke (2007)
(see also  Markoff, Nowak \& Wilms 2005).

Monitoring of the SgrA* centroid position has the 
potential to place significant constraints upon the existence and 
morphology of inhomogeneities in a super-massive black hole accretion flow.
Reid et al. (2008) present measurements with the VLBA of the variability in the
centroid position of SgrA* relative to a background quasar at 7~mm
wavelength.  They find an average centroid wander of $71\pm 45\,\mu as$ for
timescales between 50 and  $100\,\min$ and $113\pm50\,\mu as$ for timescales
between 100 and $200\,\min$, with no secular trend. 
For these particular observations 
highly asymmetric flux density distributions
can be ruled out - as they would result from a hot spot with orbital 
radii above $15\,G M_{\rm Sgr A*}/c^2$=7.5~R$_s$ and a $>$30\% contribution to the total 7~mm flux. 
Structural variations at smaller radii or lower flux density levels
remain unconstrained. 

The velocity estimates are derived under the assumption of tangled fields 
(van der Laan 1966) in the expanding components. 
If the expansion, however, takes place in a partially aligned field
- e.g. along an outflow - then the expansion is hampered in directions perpendicular to the
field lines. The component will stay more confined and 
lower values for the expansion velocity will be derived.
An expansion of source components through shearing due to differential
rotation within the accretion disk may explain the low expansion velocities.
The recent theoretical approach of hot spot evolution due to shearing is
highlighted in Eckart et al. (2008a) and Zamaninasab et al. (2008; see also Pech\'a\v{c}ek et al., 2008).
A model that explicitly solves for the relativistic hydrodynamics
and includes low expansion speeds, with reference to SgrA*
and to the work by 
Hawley \& Balbus (1991, 1998)
was recently suggested by Yuan et al. (2008).
In their model expansion velocities of less than 0.01c close to 
the accretion disk are explained. 

From the 340~GHz light curves (see also Eckart et al. 2006a) it appears that 
highly accelerated expansion of the source components is unlikely.
Highly accelerated  or decelerated expansion would result in sharp drops or rises in the 
light curves. For Fig.~\ref{Fig:speed} we can deduce that a low velocity expansion would first
result in high flare flux levels for a few hours after the initial event. A change to
a significantly higher speed would then result in a sudden drop.
Similarly for strongly decelerated flares we would first expect a fast drop to low flux density values
followed be a longer lasting decay.


\section{Summary and discussion}
We have presented new model results for the July 7, 2004 flare.
We show that the data can successfully be explained by a
combination of a SSC and an adiabatic expansion model.
Based on this interpretation we have observed the emission of the
synchronous NIR/X-ray flares
of the compact components ($\alpha$,$\beta$) with cutoff frequencies in 
the THz domain delayed by 
1-2 hours in the 340~GHz and by 3-9 hours in the 43~GHz bands. 
The delay  times of the more extended source components with 
lower cutoff frequencies 
are less than an hour to the 340~GHz and 4-7 hours to 43~GHz bands. 
We can therefore identify this flare as one with the broadest 
coverage across the electromagnetic spectrum:
from the X-ray, through the NIR, sub-mm to mm-wavelength domain.
Other flare events with a broad frequency coverage have been 
reported by
(Yusef-Zadeh et al. 2006b; Marrone et al. 2008, Eckart et al. 2008b).

The modeling suggests the presence of two separate flares $\phi$3 and $\phi$4,
that are needed to explain the variable flux at 340~GHz and 43~GHz.
Source component $\beta$ is responsible for fitting the 
decaying flank of the 340 GHz data and
component $\alpha$ accounts in combination with $\zeta$ and $\epsilon$ 
for the overall slope and the first part of the VLA data.
Our modeling also requires the low adiabatic expansion velocity  
that describes other flares from SgrA*
(Yusef-Zadeh et al. 2007, 2008, Marrone et al. 2008, Eckart et al. 200b).  

Given that there is considerable structure in the NIR/X-ray light curves
that can be linked to flare activity at sub-mm/mm wavelengths,
it appears difficult to derive an estimate of the 
expansion velocity and power-law index $p$ of the relativistic electron distribution from
an analysis of the flare profiles alone. 
Modeling the flares through an SSC formalism coupled with adiabatic expansion shows the
importance of simultaneous NIR/X-ray measurements that preceed the radio measurements.

The expansion velocities are lower than the orbital velocities for 
flare components introduced from orbiting spot models and accretion disk
models and will not allow material to leave the immediate vicinity of the
massive black hole at the position of SgrA*.
The observed sub-mm/mm flare emission therefore can be 
explained as an expansion within the occasionally existing 
accretion disk around SgrA*
or a flaring of the disk corona.
Alternatively, the expanding source components may also have a
bulk velocity that can be substantially higher than the
observed expansion velocity. 
From the small and not strongly variable VLBI source sizes as well as the
typical flare length of 2 hours, we find an upper limit on the bulk velocity of
0.07~c.
The non-detection of a bright jet or variable source sizes or positions 
indicates that an expansion into a short jet (Eckart et al. 2006b, 2008ab) or 
an occasional accretion disk (Eckart et al. 2004) is the most likely explanation of the observed
sub-mm/mm light curves.

\begin{acknowledgements}
This work was supported in part by the Deutsche Forschungsgemeinschaft
(DFG) via grant SFB 494, the Max Planck Society through 
the International Max Planck Research School, as well as 
special funds through the University of Cologne.
\emph{Chandra} research is supported by NASA grants
NAS8-00128, NAS8-38252, GO2-3115B, and G05-6093X.  We are grateful to all members
of the NAOS/CONICA and the ESO PARANAL team.  
Macarena Garc\'{\i}a-Mar\'{\i}n is supported by the German federal department for
education and research (BMBF) under the project numbers: 50OS0502 \& 50OS0801.
M. Zamaninasab, D. Kunneriath,
 are members of the International Max Planck Research School (IMPRS) for
Astronomy and Astrophysics at the MPIfR and the Universities of
Bonn and Cologne. R. Sch\"odel acknowledges support by the Ram\'on y Cajal
program by the Ministerio de Ciencia e Innovaci\'on of the
government of Spain.

\end{acknowledgements}

\end{document}